\documentclass{article}

\usepackage{arxiv}
\usepackage{parskip}
\usepackage{wrapfig}

\usepackage[utf8]{inputenc} 
\usepackage[T1]{fontenc}    
\usepackage{hyperref}       
\usepackage{url}            
\usepackage{booktabs}       
\usepackage{amsfonts}       
\usepackage{nicefrac}       
\usepackage{microtype}      
\usepackage{xcolor}         

\usepackage{caption}        
\usepackage{subcaption}     
\usepackage{graphicx}

\title{
CMIP X-MOS: Improving Climate Models with Extreme Model Output Statistics}

\author{
  Vsevolod Morozov, Artem Galliamov, Aleksandr Lukashevich, \\  
  {\bf Antonina Kurdukova, and Yury Maximov}\\[2ex]
  \texttt{Contact email: lukashevich@iccda.io} \\[3ex]
}

\renewcommand{\shorttitle}{CMIP X-MOS: IMPROVING CLIMATE MODELS}

\hypersetup{
pdftitle={A template for the arxiv style},
pdfsubject={q-bio.NC, q-bio.QM},
pdfauthor={David S.~Hippocampus, Elias D.~Striatum},
pdfkeywords={First keyword, Second keyword, More},
}

\begin{document}
\maketitle

\begin{abstract}
	Climate models are essential for assessing the impact of greenhouse gas emissions on our changing climate and the resulting increase in the frequency and severity of natural disasters. Despite the widespread acceptance of climate models produced by the Coupled Model Intercomparison Project (CMIP), they still face challenges in accurately predicting climate extremes, which pose most significant threats to both people and the environment. To address this limitation and improve predictions of natural disaster risks, we introduce Extreme Model Output Statistics (X-MOS). This approach utilizes deep regression techniques to precisely map CMIP model outputs to real measurements obtained from weather stations, which results in a more accurate analysis of the XXI climate extremes. In contrast to previous research, our study places a strong emphasis on enhancing the estimation of the tails of future climate parameter distributions. The latter supports  decision-makers, enabling them to better assess climate-related risks across the globe. 
\end{abstract}


\section{Introduction}
Numerous studies~\cite{field2012managing,parmesan2022climate} have unequivocally demonstrated the connection between climate change and the escalating frequency of extreme weather events, including floods, heatwaves, and intense precipitation . These events pose substantial risks to both lives and assets~\cite{larsen2015economic}. The accurate modeling of extreme tails is paramount for assessing the potential damages stemming from these climate extremes~\cite{abramov2023advancing,mozikov2023accessing,shevchenko2023climate}. A comprehensive understanding of long-term climate behavior spanning decades is of paramount importance in shaping policies aimed at mitigating the impacts of such extreme events.

Among the most notable extreme weather events are extreme precipitation~\cite{zeder2020observed}, heatwaves~\cite{frich2002observed}, and extreme wind speeds~\cite{wallbrink2009historical}. It is worth noting that, since 1980, the United States has borne witness to an astounding 332 natural disasters with an economic impact surpassing one billion dollars. In a striking statistic, 2021 alone saw a staggering \$150 billion increase in overall economic losses.

Contemporary Global Circulation Models (GCMs) have made significant advancements in representing extremes; however, they still grapple with regional and climate-variable specific limitations \cite{adeyeri2022trend}. Furthermore, the computational demands associated with GCMs, such as ensembles in CMIP6~\cite{eyring2016overview}, ERA5 reanalysis~\cite{hersbach2020era5}, or various ensembles aimed at a better statistical analysis present formidable challenges~\cite{vannitsem2018statistical}. These limitations impede progress toward addressing Climate Action, as outlined in the Umited Nations Sustainable Development Goals~\cite{katila2019sustainable}.

Several papers have proposed various statistical postprocessing methods aimed at quantifying and mitigating errors in GCMs~\cite{li2017review,vannitsem2018statistical}. One common approach is Model Output Statistics (MOS), which involves mapping GCM data to observations~\cite{maraun_widmann_2018}. For example, \cite{sa2020multi}  applied Support Vector Machine and Random Forest algorithms to create a MOS mapping for multiple variables on Borneo Island, analyzing rainfall changes up to 2100 using CMIP ensembles. Similarly, \cite{petetin2022model} employed MOS methods based on quantile mapping and gradient boosting to address Copernicus Atmosphere Monitoring Service forecasts of ozone ($O_3$). Additionally, Steininger et al.~\cite{steininger2023convmos} utilized a Convolutional Neural Network (CNN)-based MOS approach to map simulated precipitation from the RCM REMO model~\cite{jacob2001comprehensive} to E-OBS v19 observations~\cite{haylock2008european} in Europe. In a different context, \cite{chen2020model} employed a 3D CNN to map European Centre for Medium-Range Weather Forecasts (ECMWF) data to temperature measurements at the Tianjin station. Some studies~\cite{chen2022generative} harnessed generative models, training a network on ECMWF data and German weather station measurements of temperature and wind speed.

While these methods have proven effective, they share limitations such as a lack of global applicability, an incomplete coverage of weather extremes, and a lack of long-term analysis. To address these shortcomings, we introduce X-MOS, an eXtreme Model Output Statistics method. X-MOS learns a transfer mapping from CMIP6 GCM outputs to Global Summary of the Day (GSOD) data~\cite{gsod23}, combining global weather station observations. X-MOS employs an attention-based regression approach with a specific emphasis on modeling distribution tails.

\begin{figure}
    \centering
    \includegraphics[width=\textwidth]{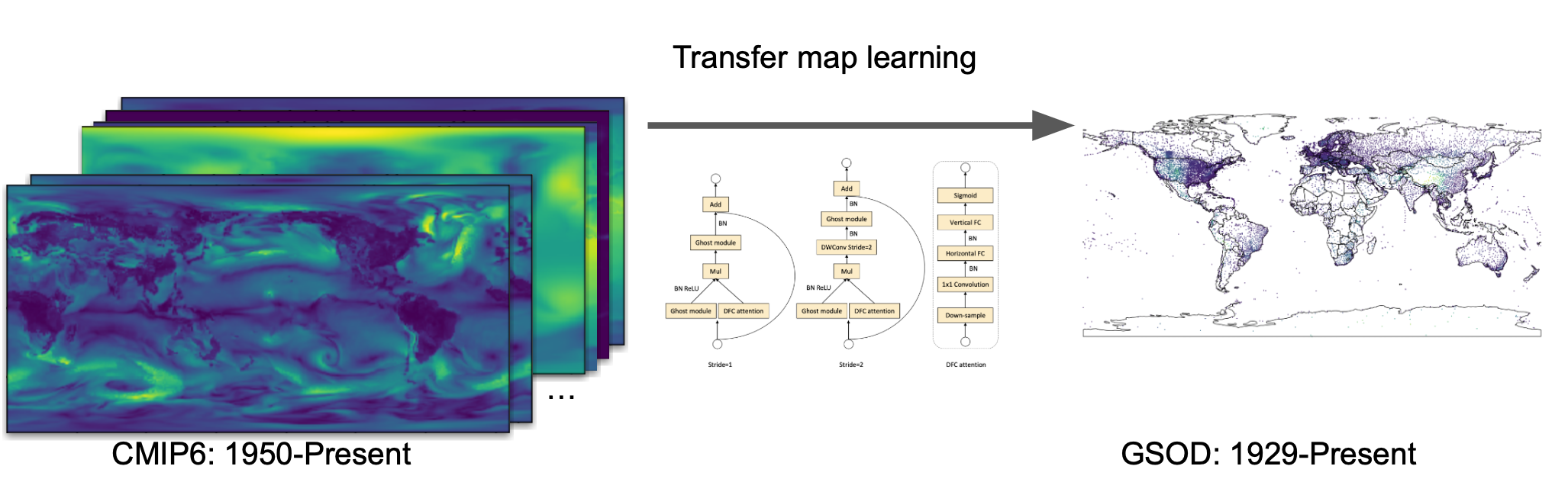}
    \caption{The X-MOS approach: GhostNetV2 is used as MOS base utility illustrated in~\cite{tang2022ghostnetv2}.}
    \label{fig:scheme}
\end{figure}

In this paper, we proved the ability of the X-MOS approach to accurately map CMIP6 GCM outputs to actual observations of temperature and wind speed. We achieve this by comparing the quantiles generated by X-MOS with those from E-OBS v27 data\cite{hersbach2020era5} for Europe and provide test metrics across the globe. Figure~\ref{fig:scheme} provides an overview of our proposed approach. Our results unequivocally demonstrate that X-MOS significantly enhances the representation of extreme values.

\section{Dataset}
\paragraph{Climate models}

This study uses several climate models, including a historical dataset and one based on the Representative Concentration Pathway (RCP) 4.5 based Shared Socio-economic Pathway (SSP) \cite{bauer2017shared}. Specifically, we utilize parts of the MRI-ACGM-3.2-H climate model from CMIP6 \cite{https://doi.org/10.22033/ESGF/CMIP6.10974,https://doi.org/10.22033/ESGF/CMIP6.10972}. The CMIP model outputs mentioned cover the entire globe with a uniform grid resolution of $0.5625\times0.5625$ for MRI-ACGM-3.2-H. In the case of E-OBS, the data grid covers only Europe, with a maximum resolution of $0.11\times0.11$ degrees. Both E-OBS v27 and MRI-ACGM-3.2-H possess a native daily temporal resolution. We have utilized the following climate variables as predictors: near-surface average wind speed~[m/s], near-surface maximum wind speed~[m/s], pressure at sea level~[Pa], and average minimum and maximum temperatures~[K].

\paragraph{Observations}
For our observations, we rely on the GSOD dataset \cite{gsod23}, which compiles daily climate measurements from over 9000 global weather stations. This dataset spans from 1929 to the present day, with a notable improvement in data quality from the 1990s onward. Daily measurements encompass mean values of temperature, dew point temperature, sea-level pressure, station pressure, visibility, wind speed, maximum/minimum temperature, maximum sustained wind speed, maximum gust, precipitation, snow depth, and weather indicators.

Furthermore, we employ E-OBS v27 \cite{Cornes2018} exclusively for testing purposes. To calculate metrics, we bilinearly interpolate both CMIP6 output and model outputs to match the grid of E-OBS v27.
In the CMIP6 output data, we focus on maximum and average surface wind speed, maximum and minimum temperature, and sea-level pressure. In the case of E-OBS v27, we concentrate on maximum and minimum temperatures and average wind speed.

\section{Approach}

This study seeks to enhance the accuracy of extreme weather representation and statistics through the year 2100. General Circulation Models (GCMs) in the CMIP family project climate conditions up to 2100 but suffer from systematic errors due to the intricate nature of modeling chaotic climate systems \cite{vannitsem2018statistical, adeyeri2022trend}. The coarse resolution and ensemble averaging in CMIP-type GCMs pose challenges when it comes to representing extreme values \cite{kim2020evaluation}. Furthermore, these models exhibit varying levels of accuracy across different regions \cite{zhu2020does} and are dependent on the institutions that publish them.

To address these issues, we employ the MOS approach with a particular focus on extreme values. Our model is trained to regress quantiles, enabling precise statistical postprocessing of CMIP-type GCMs, especially in the context of extremes and the improvement of climate projections up to 2100.
We tackle the above by using the MOS approach, concentrating on extreme values. We train our model to regress quantiles, enabling precise statistical postprocessing of CMIP-type GCMs, particularly in addressing extremes and improving climate projections up to 2100.

\paragraph{Target data}
To generate the target data, we align meteorological stations from the GSOD dataset with the geospatial grid from CMIP data, creating corresponding input data tensors. For weather stations located within a particular grid cell, we aggregate measurements by calculating the median over one day. The training target comprises several quantiles of weather measurement distributions within a 28-day time window. 

\paragraph{Addressing Extremes}
To place emphasis on extreme values, we employ regression to estimate the $0.95, 0.85, 0.70, 0.50, 0.25, 0.15,$ and $0.05$ quantiles. Our model is trained using a weighted Mean Squared Error (MSE) loss function, with higher weights assigned to the extreme quantiles, specifically $0.95$ and $0.05$. The remaining five values serve the purpose of model regularization and provide additional information about the distribution of target values.

\paragraph{Input data}
Recognizing the dynamic nature of climate and its local dependencies, we incorporate neighboring points from the climate data grid in both space and time. Each point includes data on wind gust, surface wind speed, temperature, and sea-level pressure. To perform statistical postprocessing of a CMIP model up to 2100, we leverage non-causality by considering the point's temporal vicinity both backward and forward in time.
For the target values that we predict at time $t$, we consider approximately $3900$ km in each direction, forming a square area of $3900\times3900$ km or $135\times135$ km grid points. Additionally, we include the time span from $t-27$ to $t+27$, encompassing the CMIP simulation horizon. Our goal is to predict the maximum target value within this spatio-temporal block of $55\times135\times135$. Climate variables add another dimension to the input data tensor. For instance, when utilizing six climate variables, the resulting input tensor has a shape of $6\times55\times135\times135$, where channel count, temporal, and spatial sizes are hyperparameters, and these values have proven optimal for our experiments.

\paragraph{Architechture} To accelerate computation, we separate the spatio-channel and temporal components. For extracting spatio-channel features, we utilize a network based on GhostNetV2, a deep convolutional network with DFC attention \cite{tang2022ghostnetv2}. We selected this architecture for its capacity to capture long spatial dependencies while maintaining computational efficiency. More complex networks did not yield improved results, so we opted for the fastest one available. To handle temporal aggregation of spatio-channel features, we employ a self-attention mechanism \cite{1706.03762}, which translates the data into quantiles of the target distribution.

\section{Experiment}

We assess the performance of the proposed neural network in comparison to the following baselines: direct quantile calculation (CMIP) from CMIP data and Linear Regression (LR), which employs solely temporal data from a given grid point. To enhance data quality, we exclude data from sea buoys, stations with irregular measurements, and stations reporting anomalous measurements. 

\paragraph{Experimental protocol}
We train our model to predict quantiles of measurements using data spanning from 1950 to 2022. The training dataset is divided into two parts: data up to 2008 is used for training, and data from 2009 to 2022 is reserved for testing. Additionally, we introduce a time window of 55 days to prevent data leakage. Furthermore, the features are standardized based on the training data, ensuring a zero mean and unit standard deviation. We explored various architectures, including 3D CNNs, ConvLSTM, and various 2D feature extractors. Additionally, we experimented with different spatiotemporal block dimensions, ranging from 1 to 60 pixels and from 1 to 71 pixels. The dimensions were selected based on performance as further increase provided only mild gains.

\paragraph{Metrics}
For model evaluation, we employ Mean Absolute Error (MAE) for different quantiles to assess the overall fit of the data distribution. Additionally, we use Average Precision to evaluate our model's ability to predict months with extreme weather conditions. The Average Precision is computed using a threshold of 40°C for temperature and 20 m/s for wind speed. In this calculation, ground truth values are transformed into binary values (0 and 1) using the specified threshold, and model predictions are scaled and clipped to the (0, 1) range.
All models were trained on an A100 with 80 GB of GPU memory. Training ranged from 1 to 8 epochs, with each epoch taking between 1 to 25 hours, depending on the architecture.

\section{Results}
Table \ref{tab:metrics} demonstrates that the proposed X-MOS approach outperforms the baseline algorithms in tail modeling. The notable discrepancy between the wind and temperature scores can be attributed to the inherently more chaotic nature of wind, making it inherently more challenging to predict accurately.

\begin{table}
  \centering
  \begin{tabular}{llll}
    Model / Metric     & $\texttt{MAE}_{0.95}$  & $\texttt{MAE}_\texttt{mean}$ & \texttt{AP} \\
    \midrule
    CMIP temp        & 5.71  & 5.10  & 0.89   \\
    LR temp         & 7.00  & 5.92  & 0.94   \\
    X-MOS temp    & \bf{2.12} & \bf{2.03} & \bf{0.97}  \\  \hline 
    CMIP wind        & 3.57  & 3.00  & 0.28   \\
    LR wind         & 4.28  & 2.44  & 0.28   \\
    X-MOS wind     & \bf{3.10} & \bf{1.66} & \bf{0.57}  \\
    \bottomrule
  \end{tabular}
  \caption{Experimental results on the OBS data; the metrics are means over the (globe) test set}
  \label{tab:metrics}
\end{table}

\begin{figure}
     \centering
     \begin{subfigure}[b]{0.49\textwidth}
         \centering
        \includegraphics[width=0.9\textwidth]{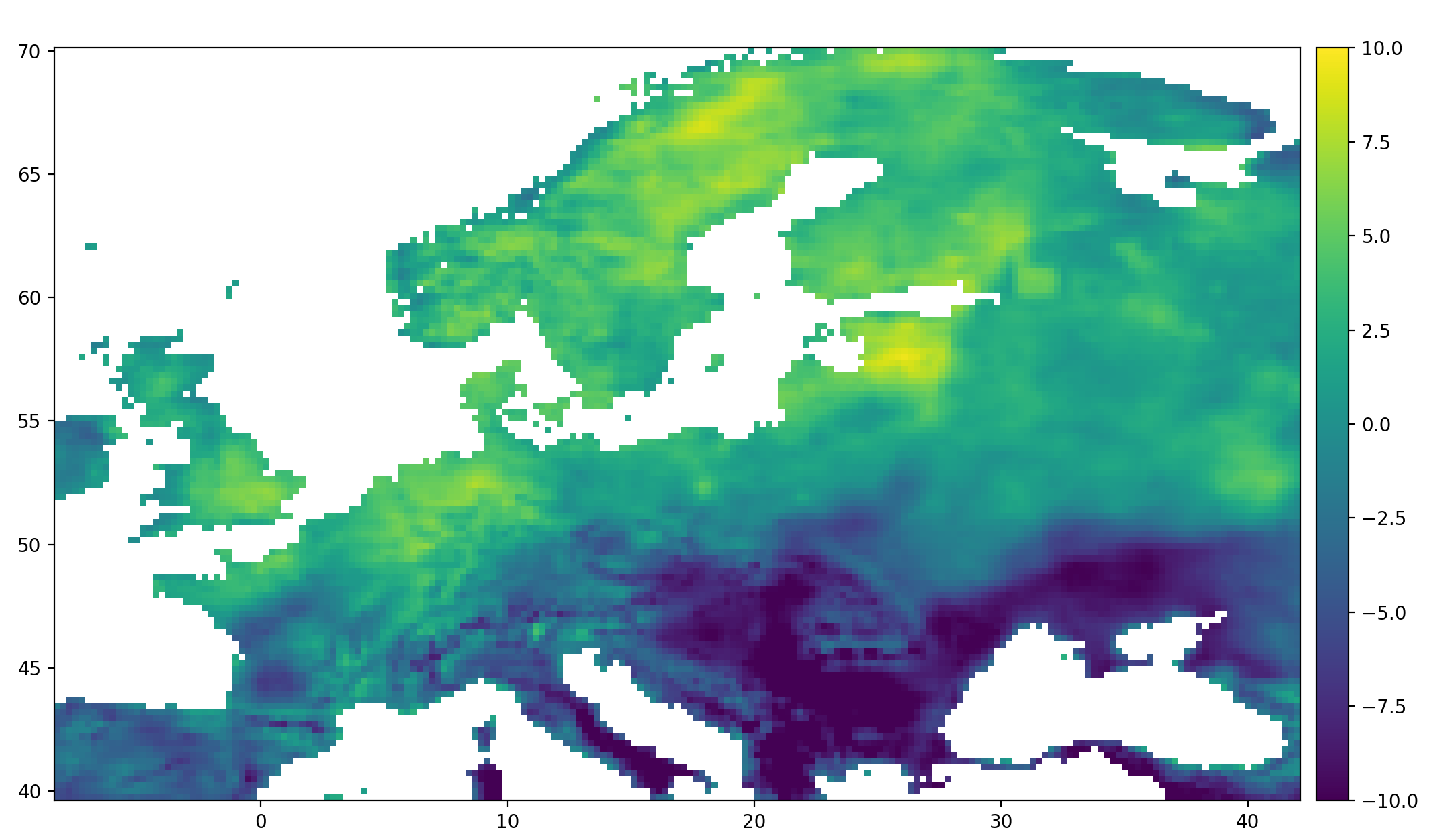}
     \end{subfigure}
     \begin{subfigure}[b]{0.49\textwidth}
         \centering
         \includegraphics[width=0.9\textwidth]{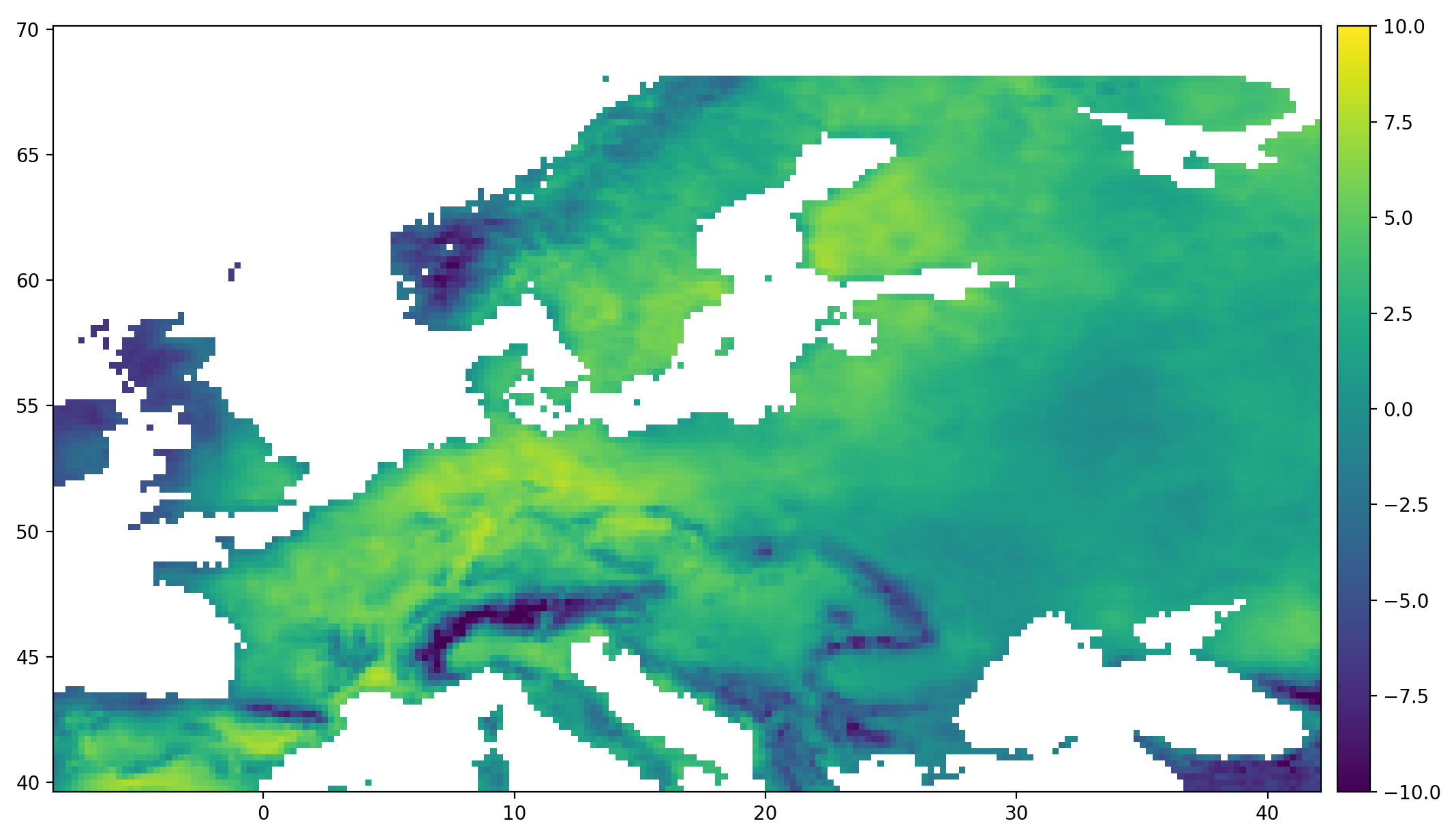}
         \label{fig:visual2}
     \end{subfigure}
     \caption{(Left) The difference between $0.95$ quantile tx variable of E-OBS v27 and interpolated $0.95$ quantile of CMIP6 tasmax, $2018.08.03\pm14$ days. 
     (Right) The difference between $0.95$ quantile tx variable of E-OBS v27 and interpolated $0.95$ quantile of X-MOS $2018.08.03\pm14$ days.}
     \label{fig:visual}
\end{figure}

Fig.~\ref{fig:visual} illustrates the improvement in the prediction of extreme temperature values across Europe during the 2018 heatwave. The Mean Absolute Error (MAE) calculated for this region is $3.21$ for CMIP6 tasmax and $1.72$ for the X-MOS. This comparison clearly indicates that the statistical postprocessing conducted with X-MOS offers a superior representation of climate extremes. 

\section{Conclusion}
In this paper we introduced the X-MOS postprocessing model within the Model Output Statistics (MOS) framework, designed to efficiently and precisely adjust the tail distributions of CMIP models using real measurements. Through the utilization of Convolutional Neural Networks (CNNs) and an attention mechanism, we have demonstrated the model's effectiveness through numerical evaluation. Importantly, this approach is applicable to any CMIP-type General Circulation Model (GCM).

The improvements in tail distributions of these climate models empower policy and decision-makers with more accurate data for future climate action planning. While our approach generally outperforms the standard methods considered in this study, further comparisons with advanced models are warranted in future research.

\bibliographystyle{abbrv}
\bibliography{biblio}

\end{document}